\documentclass[useAMS,usenatbib]{mn2e}
\usepackage{graphicx}
\usepackage{natbib}
\usepackage{amsmath}
\usepackage{float}
\usepackage{rotating}
\bibpunct{(}{)}{;}{a}{}{,}

\newcommand{\ie}{$i.e.,\;$}
\newcommand{\eg}{$e.g.,\;$}

\newcommand{\cf}{$cf.,\;$}

\title[]{{\em Suzaku} X-ray spectral study of the Compton-thick Seyfert galaxy NGC~5135}
\author[Veeresh Singh, Guido Risaliti, Valentina Braito and Prajval Shastri]{Veeresh Singh$^{1,2}$\thanks{E-mail: veeresh@iiap.res.in}, Guido Risaliti$^{3,4}$, Valentina Braito$^{5}$ and Prajval Shastri$^{1}$ \\
$^{1}$Indian Institute of Astrophysics, Bangalore 560034, India\\
$^{2}$Department of Physics, University of Calicut, Calicut 673635, India\\ 
$^{3}$INAF-Osservatorio di Arcetri, Largo E. Fermi 5, I-50125 Firenze, Italy\\ 
$^{4}$Harvard-Smithsonian Center for Astrophysics, 60 Garden St. Cambridge, MA 02138, USA\\
$^{5}$Department of Physics and Astronomy, University of Leicester, University Road, Leicester LE1 7RH, UK}

\begin{document}

\date{Accepted xxxx xxxxxx xx; Received xxxx xxxxxx xx; in original form xxxx xxxxxx xx}

\pagerange{\pageref{firstpage}--\pageref{lastpage}} \pubyear{2010}

\maketitle

\label{firstpage}

\begin{abstract}
We present the 0.5 - 50 keV {\em Suzaku} broad-band X-ray spectral study of the Compton-thick AGN in NGC~5135. 
The {\em Suzaku} observation provides the first detection of NGC 5135 above 10 keV that allowed us, for the first time, 
to estimate the absorbing column density, 
the intrinsic X-ray luminosity, the strength of the reflection component and the viewing angle of the torus for this AGN. 
The 0.5~-~10~keV spectrum of NGC~5135 is characterized by the standard components for a Compton-thick source: 
a scattered continuum, a prominent Fe K$\alpha$ emission line (EW $\sim$~2.1~keV) and a soft excess. 
At higher energies (E $>$~10~keV) the intrinsic AGN continuum shows up, implying an absorbing column density of the order 
of $\sim$ 2.5 $\times$ 10$^{24}$~cm$^{-2}$ and the intrinsic 2.0 - 10 keV X-ray luminosity of $\sim$ 1.8 $\times$ 10$^{43}$ erg s$^{-1}$. 
Assuming a toroidal geometry of the reprocessing material we show that an edge-on view of 
the obscuring torus is preferred in this source.

\end{abstract}

\begin{keywords}
galaxies: active, galaxies: individual: NGC 5135, X-rays: galaxies
\end{keywords}

\section{Introduction}
Seyfert galaxies host active galactic nuclei (AGN) powered by accretion on to supermassive black hole.
The luminous accreting engine is supposedly surrounded by obscuring material having a torus-like geometry 
\citep{Ant93}. When the obscuring torus intercepts the observer's line-of-sight 
({\ie}type~2 AGN), the absorbing column density (N$_{\rm H}$)
for X-ray photons originating from AGN is typically higher than 10$^{22}$~cm$^{-2}$ \citep{Cappi06}.
In mildly Compton-thick AGN (N$_{\rm H}$ $\sim$ a few times of 10$^{24}$~cm$^{-2}$), the primary emission is strongly 
suppressed below 10 keV and is seen only above 10 keV, while in heavily Compton-thick AGN 
(N$_{\rm H}$ $\geq$ 10$^{25}$~cm$^{-2}$) the primary emission is strongly depressed even at energies above 10 keV due to Compton down-scattering \citep{Matt2000}.
In {\em Chandra}, {\em XMM-Newton} and earlier observations sensitive only up to 10~keV, Compton-thick AGNs have been 
identified only by using diagnostic properties such as high equivalent width (EW) of Fe K$\alpha$ line ($\geq$~1.0~keV, \citet{Bassani99}), 
low flux ratio of hard X-ray (2.0~-~10.0~keV) to [OIII] $\lambda$5007$\mbox {\AA}$ line emission \citep{Maiolino98} and there are 
no measurements of the true value of the absorbing column density.
There is evidence that obscured AGNs are much more numerous than unobscured AGNs, 
both in the local universe and at intermediate to high redshifts \citep{Risaliti99,Hasinger08}.
In recent years, observations from INTEGRAL, {\em Swift}, and {\em Suzaku} 
have increased the number of Compton-thick sources detected in hard X-ray \citep{DellaCeca08,Beckmann09,Tueller08,Severgnini11}. 
However, the number of Compton-thick sources whose broad-band spectra are analyzed in detail is still limited.
In this paper we present the {\em Suzaku} broad-band X-ray spectral properties of NGC~5135, which is one of 
the brightest known Compton-thick Seyfert type 2 galaxies, with no previous X-ray observations above 10~keV. 
NGC 5135 has not been detected above 10 keV by any of the hard X-ray detectors, {\eg}{\em BeppoSAX}-PDS, {\em Swift}-BAT, {\em INTEGRAL}-IBIS, 
other than {\em Suzaku} HXD-PIN.
NGC~5135 is a relatively nearby (redshift z $\simeq$ 0.014) galaxy and 
optically classified as a Seyfert type~2 on the basis of emission line ratios \citep{Phillips83}.
IR and UV studies have shown that NGC~5135 also contains a powerful compact nuclear starburst \citep{Bedregal09,Gonz98}.
The higher spatial resolution of {\em Chandra} ($\sim$ 0.5$^{\prime\prime}$) enables isolation of the AGN and starburst emissions, and has shown that 
the AGN in NGC~5135 is completely obscured by a column density N$_{\rm H}$ $\geq$ 10$^{24}$ cm$^{-2}$ \citep{Levenson04}. 
The Compton-thick obscuration around AGN in NGC 5135 is also inferred from ASCA observations \citep{Turner97}. 
\cite{Fukazawa11} used the {\em Suzaku} data and report the presence of high absorbing column density (N$_{\rm H}$) $\sim$ 2.0 $\times$ 10$^{24}$ cm$^{-2}$ and 
a strong (EW $\sim$ 1.6 keV) Fe K${\alpha}$ emission line in NGC 5135. The primary aim of \cite{Fukazawa11} is to study the properties of 
the Fe K line features and its correlation with the absorbing column density and AGN luminosity for a sample of Seyfert galaxies. 
In this paper, we discuss the detailed broad-band X-ray spectral analysis of NGC 5135 using the same {\em Suzaku} data, with the aim to 
characterize the reprocessing material around the AGN.
\section[]{Observations and Data Reduction}
{\em Suzaku} observed NGC~5135 on 2007 July 03 (observation ID 702005010) with an exposure time of $\sim$ 52.5 ks.
{\em Suzaku} \citep{Mitsuda07} carries four X-ray telescopes (XRTs: \cite{Serlemitsos07}) with X-ray CCD cameras (XIS) at their focal-planes. 
The XISs are sensitive to 0.2~-~12.0~keV energy band with 18$^{\prime}$$\times$18$^{\prime}$ field-of-view. 
Among the four XIS CCDs, three (XIS 0, 2, and 3) are front-illuminated (FI) and one (XIS1) is back-illuminated (BI) \citep{Koyama07}. 
There are no observations with XIS2, due to a malfunction in November 2006.
{\em Suzaku} also has a non-imaging hard X-ray detector (HXD: \cite{Takahashi07}) which has two types of detectors, the 
PIN and the GSO, which are sensitive in 10~-~700~keV energy band. \\
NGC~5135 was detected by XIS 0, 1, 3 and HXD-PIN.
We obtained the XIS and HXD event files and reduced these by following the standard
procedure described in the {\em Suzaku} reduction guide\footnote{http://heasarc.gsfc.nasa.gov/docs/suzaku/analysis/abc/}, and 
using the most recent calibration files.
For the low-energy instruments (XISs), the source spectra were extracted from
a circular region of 2.4$^{\prime}$ radius centered on the source. 
The background spectra were extracted from two circular regions of 2.2$^{\prime}$ radius offset from the source and calibration sources.
The XIS response (rmfs) and ancillary response (arfs) files were produced with 
the ftools tasks {\it xisrmfgen} and {\it xissimarfgen}, respectively, and 
the latest calibration files were used.
The spectra of the two front-illuminated CCDs (XIS0 and XIS3) were merged. 
The net XIS source spectra were binned to have minimum signal-to-noise ratio (S/N) of 4 in each energy bin and ${\chi}^{2}$ statistics
has been used.
For the HXD-PIN, we used the rev2 data that include all 4 cluster units and the
best available background \citep{Fukazawa09}, which accounts for  the instrumental background
(NXB, \cite{Takahashi07}) with systematic uncertainties of  
$\sim$~1.3\% (at 1~$\sigma$). We  then
simulated a spectrum for the cosmic X-ray background counts and added  it to the  instrumental one.
Using this background NGC~5135 is detected in the 13~-~50~keV band at $\sim 7$\%  above the
background  with a net count rate  of  (2.7$\pm$0.2)$\times$10$^{-2}$ cts s$^{-1}$ 
(a total of $\sim$ 1400 net source counts have been collected), corresponding to a signal-to-noise ratio (S/N) $\sim$ 9.2.
Since the HXD-PIN is a non-imaging detector and has large field-of-view ($\sim$ 0.56 deg $\times$ 0.56 deg), 
we checked that the HXD-PIN detection of NGC 5135 is not caused and/or contaminated by any other source.
In order to do this, we searched the HEASARC/NED database for the potential hard X-ray (E $>$ 10 keV) emitting sources 
in a circular region of the radius of 17$^{\prime}$ around the NGC 5135. 
The {\em Swift}, {\em INTEGRAL} and {\em BeppoSAX} catalogs do not report any hard X-ray source detection in this area. 
The {\em Chandra}, {\em ASCA} master catalogs report only NGC 5135 as the potential hard X-ray emitting source.
During the {\em Suzaku} observations, NGC 5135 was placed at the HXD nominal pointing, therefore, 
in the spectral analysis, we used a cross-calibration constant of 1.18 between
the HXD and XIS spectra, as suggested by the {\em Suzaku}-HXD calibration team.
\section{{\em Suzaku} X-ray Spectral fit}
The 2.0 - 10 keV {\em Suzaku} data can be fitted with a flat power law ($\Gamma$ $\sim$ 1.4) and a Gaussian for 
Fe K$\alpha$ line of high EW ($\sim$ 3.0 keV), which is broadly similar to the fit reported by previous {\em Chandra} observation \citep{Levenson04}.
The flat power law and high EW of Fe K$\alpha$ line are indicative of the heavily obscured AGN. 
In order to characterize the X-ray emission from heavily absorbed AGN, 
we attempt to fit the 0.5 - 50 keV {\em Suzaku} data using physically motivated models.
\\
In order to avoid the complexity of the soft component (E $<$~2.0~keV) we first attempted to fit the 2.0 - 50 keV spectrum with a basic model 
of obscured AGN {\ie}an absorbed power law representing the AGN X-ray emission transmitted through a cold absorber and an unabsorbed power law 
for the scattered component. This simple model gives ${\chi}^{2}$/d.o.f. $\sim$ 182/31 and leaves large residuals around 6.4 keV with an 
emission line-like shape. Adding a Gaussian profile for the Fe K${\alpha}$ at 6.4 keV improves the fit very significantly and gives the fit 
statistics ${\chi}^{2}$/d.o.f. $\sim$ 39/29, with 
N$_{\rm H}$ $\sim$ 2.5 $\times$ 10$^{24}$ cm$^{-2}$, $\Gamma$ $\sim$ 1.8 for the absorbed component, 
a steep power law ($\Gamma$ $\sim$ 2.7) accounting for the emission below 10 keV and the 
equivalent width (EW) of Fe K$\alpha$ line $\sim$ 2.1 keV. 
The Fe K$\alpha$ line is fitted with a narrow Gaussian ($\sigma$ fixed to 10 eV) and any increase in the line 
width worsen the fit. 
There are no signatures for the Fe K${\beta}$ or the ionized component of Fe K$\alpha$ line emission in the residuals and the 
addition of any Gaussian component for such lines worsen the fit statistics. 
If the photon index ($\Gamma$) of the unabsorbed power law component is fixed to the value equal to the photon index of the 
absorbed power law component, the fit statistics does not improve and this fit results $\Gamma$ $\sim$ 2.1 with little increase in 
the column density ( N$_{\rm H}$ $\sim$ 2.6 $\times$ 10$^{24}$ cm$^{-2}$), while the other parameters remain nearly unchanged.
Considering the high N$_{\rm H}$ of the absorbed component and the high EW of the Fe K$\alpha$ line as the indicative of the 
presence of reflection component, we added the {\tt PEXRAV} \citep{Magdziarz95} model component which represents the 
emission reflected from a neutral medium. 
The addition of reflection component gives ${\chi}^{2}$/d.o.f. $\sim$ 33/28 (model `M1' in Table 1) 
{\ie}improvement of ${\Delta}{\chi}^{2}$ $\sim$ 6 for 1 d.o.f. at the significance level of 90$\%$.
The photon index of the reflected component ({\tt PEXRAV}) was fixed equal to the transmitted component and the reflection scaling factor 
was fixed to -1, while all other parameters of the {\tt PEXRAV} model were kept to their default values. 
We note that an equally good fit is obtained if we fix the photon-index of the intrinsic power law to 1.9, the canonical value for Seyferts \citep{Nandra94}. 
However, as expected, a higher value of the absorbing column density (N$_{\rm H}$ $\sim$ 2.9 $\times$ 10$^{24}$ cm$^{-2}$) is required 
to account for the steeper power law.
In order to assess the importance of the reflection component, we estimated the 2.0 - 10 keV X-ray fluxes associated with the 
unabsorbed power law and the {\tt PEXRAV} (reflection) model components using the model `M1'. 
The 2.0 - 10 keV X-ray fluxes associated with the unabsorbed power law and the {\tt PEXRAV} (reflection) model components are 
$\sim$ 1.27 x 10$^{-13}$ erg cm$^{-2}$ s$^{-1}$ and $\sim$ 1.49 x 10$^{-13}$ erg cm$^{-2}$ s$^{-1}$, respectively. 
These flux values correspond to $\sim$ 28$\%$ and $\sim$ 34$\%$ of the total observed flux in 2.0 - 10 keV band, respectively, and 
suggest that both the unabsorbed power law and the reflection components contribute substantially in this band. 

\par
Since Compton scattering is significant for large column densities and high energy photons 
\citep{Yaqoob97}, we applied the scattering component using {\tt CABS} model to the primary transmitted component 
(model `M1$^{\dagger}$' in Table 1). 
However, the {\tt CABS} model does not account for the photons scattered into the line-of-sight from other directions. 
To account for this effect, we fitted the spectrum with a model (`M1$^{\ddagger}$' in Table 1) in which 
the {\tt PLCAB} model accounts for Compton scattering assuming a uniform, spherically distributed reprocessing material 
around the X-ray emitting source \citep{Yaqoob97}. 
This model also gives fit statistics (${\chi}^{2}$/d.o.f. $\sim$ 32.5/27) and parameters similar to the 
models `M1' and `M1$^{\dagger}$' ({\cf}Table 1). 
\\
We also checked the spectral fitting with a model (`M2' in Table 1) in which the reflection component is accounted by the {\tt REFLIONX} model 
\citep{Ross05}. The {\tt REFLIONX} model characterizes the reflected emission from an optically thick disc illuminated by 
radiation with a power law spectrum and produces the florescent emission lines as well as the continuum emission.
The best-fit which could fit the Fe K$\alpha$ emission line gives unusually high Fe abundance 
($\sim$ 10 times Solar), low ionization parameter ($\xi$ $\sim$ 91$^{+160}_{-78}$ {\bf erg cm s$^{-1}$}) and photon index ($\Gamma$) of $\sim$ 1.6 
with ${\chi}^{2}$/d.o.f. $\sim$ 37.6/28. The required high abundance and the low ionization parameter suggest 
that the Fe line emission is originating mainly from neutral or mildly ionized material. \\
Since the {\tt PLCAB} model is valid up to energies between 10 and 18.5 keV and for column densities up to $\sim$ 5 $\times$ 10$^{24}$~cm$^{-2}$ 
\citep{Yaqoob97}, we attempted to use the {\tt MYTORUS} model \citep{Murphy09} that accounts for the 
photoelectric absorption, Compton down-scattering as well as Fe K$\alpha$, K$\beta$ line emission and 
is valid for the 0.5 - 500 keV energy band and for column density up to 10$^{25}$ cm$^{-2}$.
This model considers an azimuthally symmetric toroidal geometry around AGN for the X-ray reprocessor and produces 
the reprocessed continuum and Fe K$\alpha$, Fe K$\beta$ lines self-consistently. 
The opening angle of the torus is fixed to 60$^{\circ}$ and the column density as well as the viewing angle are free parameters. 
The best-fit using the {\tt MYTORUS} model (`M5') renders N$_{\rm H}$ $\sim$ 2.7 $\times$ 10$^{24}$ cm$^{-2}$, $\Gamma$ $\sim$ 1.96, and 
the viewing angle of the torus $\sim$ 90$^{\circ}$, with ${\chi}^{2}$/d.o.f. $\sim$ 45/27. 
The best-fit is obtained by fixing the viewing angle to 90$^{\circ}$. If we allow it to vary, the fit statistics does not improve 
({\ie}similar ${\chi}^{2}$ value with decrease of 1 in d.o.f.) and gives viewing angle $\geq$ 88$^{\circ}$ at 90$\%$ confidence level.
We note that while using the {\tt MYTORUS} model (`M5'), an additional unabsorbed power law of photon index $\sim$ 2.0 is 
required to fit the data below 10 keV.
\\
We finally fitted the whole 0.5 - 50 keV spectrum with the above models and noted that addition of a thermal plasma model 
({\tt MEKAL} in XSPEC) of temperature 
kT $\sim$ 0.70 keV, is required to account for the emission below 2.0 keV (models `M3', `M3$^{\dagger}$', `M3$^{\ddagger}$' and `M4' in Table 1).
While fitting the 0.5~-~50~keV spectrum, we have excluded the XISs data points lying in 1.6~-~1.9~keV energy range 
because they are likely to be affected by systematic calibration uncertainties around the instrumental silicon K-edge. 
We find that the best-fit parameters obtained are similar in both the 2.0 - 50 keV and 0.5 - 50 keV spectral fittings. 
This implies that the addition or removal of the soft component (E $<$~2.0~keV) does not affect the interpretation of the hard components.
Table 1 lists the models and their resulting best-fit parameters. 
All the quoted errors are at 90$\%$ confidence level for one interesting parameter.
The best-fitted spectrum using model `M3' and the unfolded model are shown in Figure 1. 
Figure 2 shows the confidence contours for the photon index versus absorbing column density using model `M3'.
\\
\begin{table*}
\centering
\begin{minipage}{140mm}
\caption{The best-fit spectral parameters}
\begin{tabular}{@{}cccccccccc@{}}
\hline
Energy  & Model &  \multicolumn{2}{c} {\it Soft Component} &  \multicolumn{3}{c} {\it Hard Component} & \multicolumn{2}{c} {\it Fe K$\alpha$ line} & 
${\chi}^{2}$/{\it dof} \\
range  &       & kT  & ${\Gamma}_{\rm soft}$  &  N$_{\rm H}$  & ${\Gamma}_{\rm hard}$  & R & E$_{\rm line}$  & EW$_{\rm Fe}$  &    \\ 
         &       & (keV) &       & ($\times$ 10$^{24}$ cm$^{-2}$)  &       &      &  (keV)     & (keV)     &   \\ \hline
2.0 - 50.0 & M5 &  ...   &  2.04$^{+0.26}_{-0.08}$   &  2.71$^{+0.81}_{-0.75}$ &  1.96$^{+0.26}_{-0.08}$ & ... &  ... & ... &  44.9/27 \\
2.0 - 50.0 &  M1 &  ...     & 2.76$^{+0.29}_{-0.25}$  &  2.50$^{+1.28}_{-0.92}$   &  1.50$^{+1.05}_{-0.71}$  &  0.40$^{+0.58}_{-0.27}$   &  6.38$^{+0.01}_{-0.01}$   & 2.01$^{+0.44}_{-0.27}$  &  32.8/28 \\
2.0 - 50.0 & M1$^{\dagger}$ &   ...    & 2.76$^{+0.22}_{-0.18}$  & 2.51$^{+1.06}_{-1.08}$   &  1.51$^{+1.01}_{-0.66}$  & ... &  6.38$^{+0.01}_{-0.01}$   & 2.14$^{+0.36}_{-0.38}$  &  32.6/28 \\
2.0 - 50.0 &  M1$^{\ddagger}$ & ...   & 2.88$^{+0.16}_{-0.22}$  &  2.33$^{+1.03}_{-0.91}$   &  1.68$^{+0.90}_{-0.71}$  &  ... &  6.38$^{+0.01}_{-0.01}$   & 2.09$^{+0.42}_{-0.38}$  &  32.5/27 \\
2.0 - 50.0 & M2 & ...    &     2.17$^{+0.78}_{-0.52}$  & 2.34$^{+0.56}_{-0.36}$ &  1.64$^{+0.28}_{-0.16}$  &  ... & ... & ...  &  37.6/28 \\ 
0.5 - 50.0 &  M3 & 0.67$^{+0.09}_{-0.04}$  & 2.73$^{+0.19}_{-0.17}$ & 2.59$^{+1.04}_{-0.68}$ & 1.55$^{+1.09}_{-0.66}$ & 0.42$^{+1.64}_{-0.21}$  &  6.38$^{+0.01}_{-0.01}$ &  2.10$^{+0.46}_{-0.32}$  &  89.0/70  \\
0.5 - 50.0 &  M3$^{\dagger}$ & 0.73$^{+0.07}_{-0.03}$  & 2.82$^{+0.18}_{-0.18}$ & 2.74$^{+1.76}_{-1.05}$ & 1.59$^{+1.02}_{-0.69}$ & ...  &  6.38$^{+0.01}_{-0.01}$ &  2.07$^{+0.44}_{-0.38}$  & 91.4/70  \\
0.5 - 50.0 &  M3$^{\ddagger}$ & 0.67$^{+0.08}_{-0.03}$  & 2.71$^{+0.19}_{-0.18}$ & 2.30$^{+1.00}_{-0.98}$ & 1.60$^{+0.97}_{-0.79}$ & ...  &  6.38$^{+0.01}_{-0.01}$ &  2.09$^{+0.40}_{-0.28}$  &  89.0/70 \\
0.5 - 50.0 & M4 &   0.80$^{+0.06}_{-0.06}$  &  2.31$^{+0.12}_{-0.07}$   &   2.20$^{+0.36}_{-0.27}$  & 1.65$^{+0.26}_{-0.12}$   & ... & ... & ... &  107.7/72  \\ \hline 
\end{tabular}
\\ Notes:- M1:~const(pl+wabs*pl+pexrav+line), M1$^{\dagger}$:~const(pl+wabs*cabs*pl+pexrav+line), 
M1$^{\ddagger}$:~const(pl+plcabs+pexrav+line), M2:~const(pl+wabs*pl+reflionx.mod), M3:~const(mekal+pl+wabs*pl+pexrav+line), 
M3$^{\dagger}$:~const(mekal+pl+wabs*cabs*pl+pexrav+line), M3$^{\ddagger}$:~const(mekal+pl+plcabs+pexrav+line), 
M4:~const(mekal+pl+wabs*pl+reflionx.mod), M5:~{\tt MYTORUS} \citep{Murphy09}.
\end{minipage}
\end{table*}
%
\begin{figure*}
\includegraphics[angle=0,width=8.4cm]{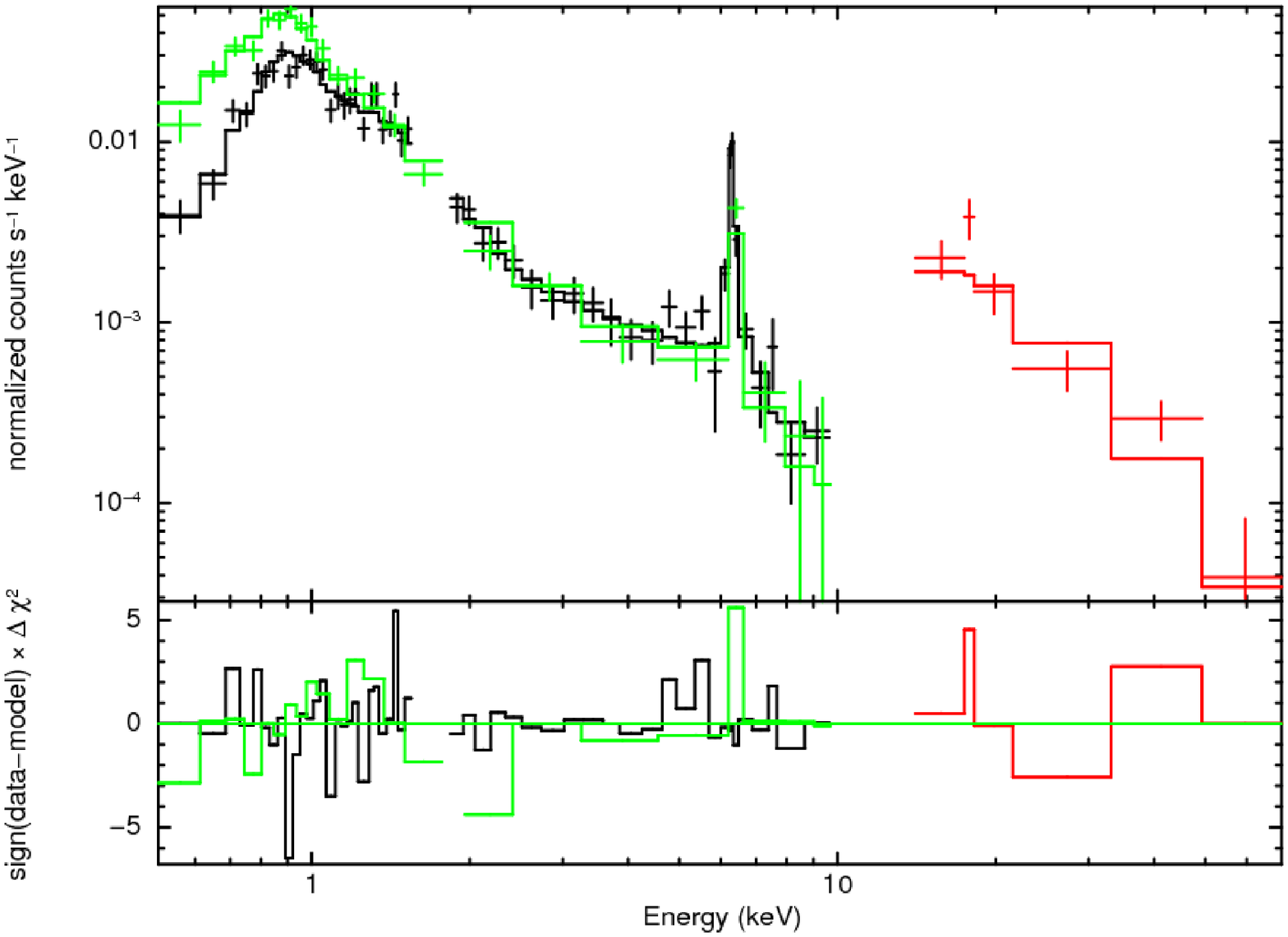}{\includegraphics[angle=0,width=8.2cm]{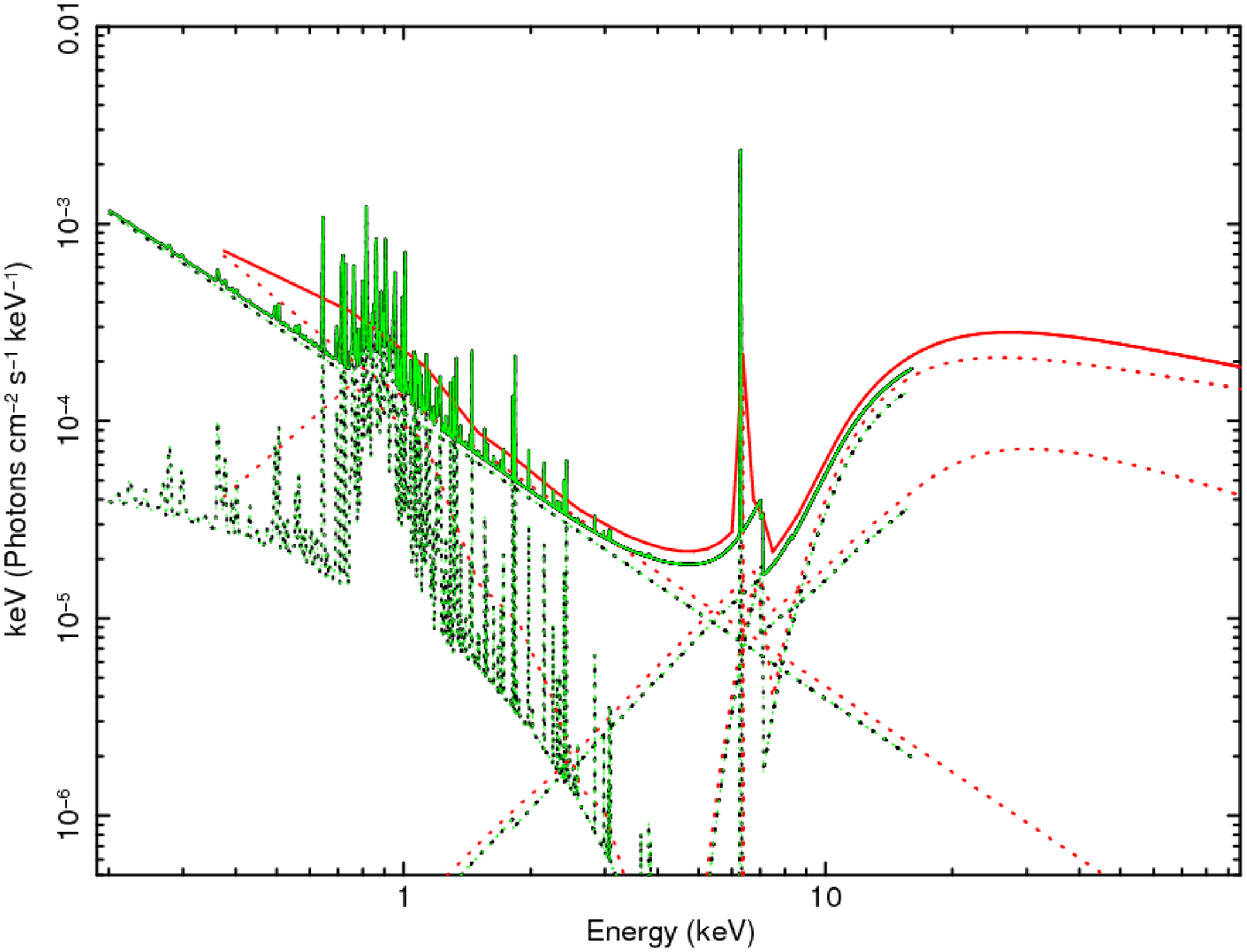}}
\caption{{\em Suzaku} 0.5~-~50~keV spectral fit using model `M3'. 
The left panel shows the spectral fit and residuals (in the bottom panel) and the right panel shows the unfolded model `M3'. 
The XIS0+3, XIS1 and HXD-PIN data points and spectral components are shown in Black, Green and Red, respectively. 
In unfolded model, the additive components are shown by dotted curves and cumulative model is shown by solid curves.}
\end{figure*}
\begin{figure}
\includegraphics[angle=0,width=8.0cm]{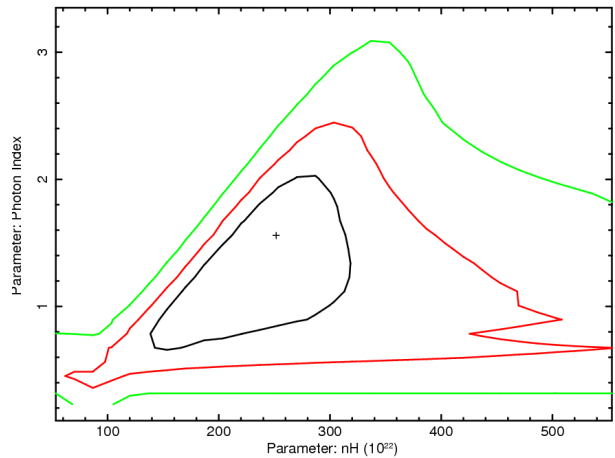}
\caption{Confidence contours for the photon index versus absorbing column density obtained from model `M3'. 
Contours at 1$\sigma$, 2$\sigma$, and 3$\sigma$ level are shown in Black, Red and Green, respectively.}
\end{figure}
\begin{table}
\caption{{\em Suzaku} X-ray fluxes and luminosities of NGC~5135}
\begin{tabular}{@{}ccccc@{}}
\hline
Energy & Model & F$_{\rm obs}$ &  L$_{\rm obs}$ & $\frac{\rm F_{\rm cor}}{\rm F_{\rm obs}}$ \\
(keV)  &   & (erg cm$^{-2}$ s$^{-1}$)  &  (erg s$^{-1}$) &     \\  \hline
0.5 - 2.0 &  M3      &  3.89 $\times$ 10$^{-13}$  & 1.66 $\times$ 10$^{41}$ & 5.0  \\
         & M3$^{\dagger}$ & 3.89 $\times$ 10$^{-13}$  & 1.66 $\times$ 10$^{41}$ &  65.9 \\
         & M3$^{\ddagger}$ &  3.89 $\times$ 10$^{-13}$ & 1.66 $\times$ 10$^{41}$ &   16.6 \\
         & M4      &  3.88 $\times$ 10$^{-13}$   &  1.66 $\times$ 10$^{41}$  &  12.3 \\
2.0 - 10 & M3               & 4.44 $\times$ 10$^{-13}$  &  1.89 $\times$ 10$^{41}$  &  11.3 \\
       & M3$^{\dagger}$   & 4.40 $\times$ 10$^{-13}$  &  1.87 $\times$ 10$^{41}$  &  124.8 \\
       & M3$^{\ddagger}$  & 4.43 $\times$ 10$^{-13}$  &  1.89 $\times$ 10$^{41}$  &   27.0 \\
       & M4               &  4.37 $\times$ 10$^{-13}$ &   1.86 $\times$ 10$^{41}$  &  20.5 \\
       & M5               &  4.17 $\times$ 10$^{-13}$ &  1.78 $\times$ 10$^{41}$ &  135.5 \\
10 - 50& M3               & 1.58 $\times$ 10$^{-11}$  &  6.67 $\times$ 10$^{42}$ &   1.2 \\
       & M3$^{\dagger}$  & 1.50 $\times$ 10$^{-11}$  & 6.39 $\times$ 10$^{42}$ &   8.4  \\
       & M3$^{\ddagger}$  &  1.50 $\times$ 10$^{-11}$   &  6.67 $\times$ 10$^{42}$  &    1.7 \\
       & M4       &  1.49 $\times$ 10$^{-11}$       & 6.35 $\times$ 10$^{42}$  &  1.3 \\
       & M5       &  1.46 $\times$ 10$^{-11}$ &  6.17 $\times$ 10$^{42}$  &  4.5 \\
Fe K$\alpha$ & M3  &  9.35 $\times$ 10$^{-14}$   &   3.98 $\times$ 10$^{40}$ &  ...     \\
\hline
\end{tabular}
Notes:- F$_{\rm obs}$: observed flux, L$_{\rm obs}$: observed luminosity, F$_{\rm cor}$: absorption corrected flux.
\end{table}

\section{Discussion}
\subsection{Soft X-ray emission} 
In our {\em Suzaku} X-ray spectral fitting, the soft X-ray emission below 2.0~keV is accounted by a 
thermal plasma model of temperature kT $\sim$~0.70~keV and a steep unabsorbed power law. 
The observed 0.5~-~2.0~keV luminosity is $\sim$ 1.66 $\times$ 10$^{41}$~erg~s$^{-1}$ ({\cf}Table~2), 
consistent with previous  {\em ASCA} and {\em Chandra} observations within the measurement uncertainties. 
The ratio of soft X-ray to far infrared emission (L$_{\rm 0.5 - 2.0~keV}$/L$_{\rm FIR}$) $\sim$ 3.0 $\times$ 10$^{-4}$ for NGC~5135 
is consistent with the typical ratio observed in starburst galaxies ({\eg}\cite{Ranalli03}), suggesting that the soft X-ray emission 
is dominated by starburst emission. 
Indeed, {\em Chandra} observation of high spatial resolution shows that the 0.5 - 2.0 keV soft X-ray emission 
from the central region is dominated by the circumnuclear starburst. 
The soft X-ray spectral components {\ie}thermal component of temperature kT $\sim$ 0.70 keV and a steep power law 
($\Gamma$ $\sim$ 2.7), in our {\em Suzaku} data are fairly consistent with the X-ray spectral properties of a spatially-resolved strong 
circumnuclear starburst region identified in {\em Chandra} observation \citep{Levenson04}. The steep unabsorbed power law is likely to 
have contribution from different individual sources, mainly X-ray binaries in the starburst region \citep{Zezas02}.
High resolution near-IR study of the central $\sim$ 2.3 kpc region in NGC 5135 has revealed the star formation knots and 
a large $\sim$ 600 pc [Si VI] 1.96$\mu$m line emission ionization cone centered on AGN \citep{Bedregal09}. 
It has been argued that SNR shocks play a dominant role in ionizing the gas in central $\sim$ 2.3 kpc region and 
photoionized emission due to AGN as well as the recent star formation have rather localized and smaller contribution \citep{Bedregal09}.
We note the fact that in general, high resolution X-ray grating spectral studies show that the soft X-ray emission in Seyfert 2s is dominated 
by recombination lines implying photoionization as the primary mechanism \citep{Kinkhabwala02,Guainazzi07,Marinucci11}. 
However, the soft X-ray emission in Seyfert 2s with intense circumnuclear starburst, is likely to be dominated by collisionally 
ionized optically thin plasma emission associated with star formation \citep{Guainazzi09}. 
Limited spatial and spectral resolution of the {\em Suzaku} XIS CCD does not allow us to quantify the AGN and starburst contribution to 
the 0.5 - 2.0 keV soft X-ray emission, which in turn, unable us to draw firm conclusions about the origin of the soft X-ray emission.
X-ray observations of high spectral and spatial resolution are required to confirm the nature of the soft X-ray emission in NGC 5135.
\\
\subsection{Hard X-ray emission} 
Our {\em Suzaku} X-ray spectral fitting shows that the 2.0 - 10 keV continuum emission has contribution from several spectral components 
{\eg}steep unabsorbed power law, thermal plasma model, reflection component, absorbed power law (see, Figure 1). However, the relative 
contribution from each individual spectral component is energy dependent. For example, the relative contribution of the unabsorbed power law 
decreases towards higher energies, while it increases for the reflection component. 
The thermal plasma model and absorbed power law components contribute only to the soft and hard part of the 2.0 – 10 keV band, respectively.
Using SFR - X-ray correlation (SFR $\simeq$ 2.0 $\times$ L$_{\rm 2.0-10~keV}$ M$_{\odot}$ yr$^{-1}$) from \cite{Ranalli03} and assuming SFR $\sim$ 
15 M$_{\odot}$ yr$^{-1}$ in NGC 5135 \citep{Bedregal09}, we noted that the estimated contribution from the circumnuclear starburst to the 2.0 - 10 keV 
X-ray luminosity is $\sim$ 7.5 x 10$^{40}$ erg s$^{-1}$, which $\sim$ 40$\%$ of the total observed 2.0 - 10 keV luminosity. 
This value is similar to the 2.0 - 10 keV non-AGN X-ray luminosity reported in {\em Chandra} observation, which could spatially resolve the 
AGN and circumnuclear starburst X-ray emission \citep{Levenson04}.
Above argument infers that nearly $\sim$ 60$\%$ of the total observed 2.0 - 10 keV X-ray emission is attributed to AGN emission, 
which is accounted by the scattered, reflected and transmitted components.
Our {\em Suzaku} (0.5 - 50 keV) broad-band X-ray spectral analysis of NGC 5135 shows the characteristics of starburst as well as 
obscured AGN and thus confirming its composite nature. 
Furthermore, the X-ray emission from circumnuclear starburst regions 
in NGC 5135 can be characterized by both thermal plasma model plus unabsorbed power law \citep{Levenson04}.
It is noted that star-forming galaxies show characteristic power law X-ray emission that could be the integrated spectrum of distinct X-ray sources 
\citep{Zezas02}. Moreover, an unabsorbed power law component representing the scattered emission of the primary AGN continuum to the line-of-sight 
is also commonly seen in obscured AGNs \citep{Braito09,Comastri10,Severgnini11}. 
With present Suzaku data we cannot spatially separate out the AGN and circumnuclear starburst emission and 
therefore it is possible that the unabsorbed power law component in our spectral fitting may have contribution from both the {\bf circumnuclear} 
starburst as well as the scattered component of AGN emission. 
It is likely that the quality of the present 2.0 - 50 keV Suzaku data do not allow us to accurately determine the relative contributions from 
starburst, scattered, reflected and absorbed AGN emission spectral components in the 2.0 - 10 keV energy band.
\par
Our {\em Suzaku} 2.0 - 50 keV broad-band spectral fitting clearly shows that the hard X-ray continuum is characterized by a 
heavily absorbed power law and a reflection component, consistent with the previous results reported by \cite{Fukazawa11}. 
The high absorbing column density (N$_{\rm H}$ $\sim$ 2.5 $\times$ 10$^{24}$ cm$^{-2}$) and 
the high ratio of the intrinsic to the observed 2.0~-~10~keV flux/luminosity (see, Table 2) infer the presence of a heavily obscured, luminous AGN. 
The ratio of the intrinsic to the observed flux/luminosity in 2.0 - 10 keV band, found for NGC 5135, is consistent with the typical 
values observed in other Compton-thick AGNs \citep{Levenson06,Brightman11}.
Given the importance of Compton scattering in absorbers with column densities of a few times of 10$^{24}$~cm$^{-2}$, 
the absorption corrected (intrinsic) flux/luminosity is model-dependent. 
The value of the intrinsic luminosity ($\frac{\rm L_{\rm 2.0~-~10~keV, int}}{\rm L_{\rm 2.0~-~10~keV, obs}}$ $\sim$ 125) obtained from the 
model `M3$^{\dagger}$' which uses the {\tt CABS} model for the scattering, can be considered as the upper limit. 
Since the {\tt CABS} model represents the Compton scattering from the matter only along the line-of-sight and 
therefore offers a small covering fraction {\ie}a single cloud along the line-of-sight which only removes photons from the beam owing to 
Compton scattering and does not consider the addition of photons scattered from other directions into the line-of-sight.
Moreover, the ratio of the intrinsic to the observed flux/luminosity also depends on the photon index of the intrinsic power law, 
for example, this ratio obtained from M3$^{\dagger}$ model is as high as $\sim$ 300, if we fix the photon index to 1.9.
The value of the intrinsic luminosity obtained from the model `M3$^{\ddagger}$' where the {\tt PLCABS} model is used for scattering 
can be considered as a lower limit, since in this scenario, the contribution from photons initially not in the line-of-sight, 
which are then scattered to the line-of-sight, is highest, and roughly compensate the loss of photons which are initially
along the line-of-sight, and that are scattered out.
Thus, the two models represent two limits: a single cloud covering ({\tt CABS}) gives an upper limit and 
full covering ({\tt PLCABS}) gives a lower limit on the intrinsic luminosity, and the true value lies somewhere in between 
depending on the covering factor. 
In the {\tt MYTORUS} model covering factor corresponds to 0.5 and the ratio of the intrinsic to the observed flux/luminosity 
($\sim$ 135) indeed lies between the two limits, if the photon index of the intrinsic power law is assumed to be 1.9.
\\ 
High obscuration around the AGN is also inferred from the flux ratio of hard (2.0~-~10~keV) X-ray to 
[OIII] $\lambda$5007{\mbox {\AA}} line emission \citep{Bassani99}, {\ie}for NGC 5135, we obtain 
log[F$_{\rm 2.0~-~10~keV,~obs}$/F$_{\rm [O III]}$] $\simeq$ -1.1, using the reddening-corrected [OIII] flux given in \cite{Singh11}. 
Furthermore, the flux ratio of the intrinsic 2.0~-~10~keV hard X-ray to [OIII] $\lambda$5007{\mbox {\AA}} line emission is
(F$_{\rm 2.0~-~10~keV,~int}$/F$_{\rm [O III]}$) $\sim$ 9.2, consistent with the mean ratio of 
observed in Seyfert 1 galaxies \citep{Heckman05}, 
supporting the notion that the obscured nucleus in NGC~5135 is intrinsically similar to a Seyfert type~1 nuclei. \\
Signature of Compton-thick matter around the AGN can be characterized by Compton reflection continuum with 
a broad hump peaking around 20~-~30~keV, which rapidly decreases at both low and high energies due to photoelectric absorption 
and Compton down-scattering, respectively. In our spectral fits the reflection strength measured as the ratio of the normalizations 
of the reflection component to the transmitted component (absorbed power law) is $\sim$ 0.4 (using model `M1'). 
We do not attempt to compare the strengths of the reflection components obtained by using the {\tt MYTORUS} model and the {\tt PEXRAV} model 
since these models consider different geometries for the reprocessor. 
The relative strength of the reflection continuum very critically depends on geometry since it is affected by 
the angle of reflection integrated over the surface of the reprocessor.
\cite{Murphy09} showed that the reflection spectrum from a Compton-thick face-on torus that subtends the same solid angle 
as a Compton-thick face-on accretion disc at the X-ray source is much weaker and it is not meaningful to compare the reflection strength 
given by the {\tt MYTORUS} model with one from the {\tt PEXRAV} model.
\par
\subsection{Geometry of the absorber}
It is important to note that the {\tt PEXRAV} as well as {\tt REFLIONX} models consider reflection from the surface of accretion disc and 
therefore, are insufficient to constrain the geometry of the toroidal-shaped obscuring material around the AGN.
We used the {\tt MYTORUS} model \citep{Murphy09} that considers the geometry of reprocessing material as an azimuthally symmetric 
obscuring torus.
This model suggests the edge-on view ({\ie}the viewing angle $\sim$ 90$^{\circ}$) of the obscuring torus. 
The absorbing column density given by the {\tt MYTORUS} model corresponds to the equatorial column density of the toroidal-shaped 
obscuring material. We note that the column densities obtained from the {\tt MYTORUS} and other models are similar within the uncertainties 
(see, Table 1) and therefore, it again confirms that the obscuring torus is oriented nearly edge-on.
We caution that in the current version of the {\tt MYTORUS} model, the torus opening angle is fixed to 60$^{\circ}$ 
while it {\bf needs} to be a varying parameter. The opening angle of 60$^{\circ}$ corresponds to covering factor of 0.5 and the solid angle 
subtended by the torus on to the X-ray emitting source to 2$\pi$.
Hints on the covering fraction of the absorber, as seen from the X-ray source, can be inferred from our fits.
It has been shown that variation in the geometry of obscuring torus, and not the iron abundance or intrinsic spectral shape, 
is required to produce EWs significantly larger than 1~keV and therefore, EW of Fe K$\alpha$ line can be used to place limits 
on the covering factor \citep{Levenson02}. \cite{Levenson02} argued that the high EW of 
Fe K$\alpha$ line ($\sim$ 2.1~keV) in NGC~5135 suggests a small opening angle ($\sim$ 20$^{\circ}$) of the obscuring torus 
giving a covering fraction of $\sim$ 90$\%$.
Our {\em Suzaku} X-ray spectral fitting gives L$_{\rm Fe}$ $\simeq$ 3.98 $\times$ 10$^{40}$ 
erg s$^{-1}$ and using \cite{Levenson06} empirical ratio of Fe K$\alpha$ line luminosity to intrinsic 2.0~-~10~keV luminosity we obtain 
L$_{\rm 2-10~keV,~int}$ $\sim$ 1.99 $\times$ 10$^{43}$ erg s$^{-1}$, {\ie}nearly 100 times the observed 2.0~-~10.0~keV luminosity. 
This value of the 2.0 - 10 keV intrinsic luminosity lies between the values from the {\tt MYTORUS} model 
(covering fraction $\sim$ 50$\%$) and the M3$^{\ddagger}$ model that uses the {\tt PLCABS} model (covering fraction $\sim$ 100$\%$).
Furthermore, we note that the intrinsic 2.0 - 10 keV X-ray luminosity obtained by using the mid-IR (12 $\mu$m) - X-ray luminosity correlation 
(L$_{\rm MIR}$ - L$_{\rm 2.0-10~keV}$) and mid-IR luminosity (L$_{\rm MIR}$) $\sim$ 1.15 $\times$ 10$^{43}$ erg s$^{-1}$ given in \cite{Gandhi09}, 
is $\sim$ 7.6 $\times$ 10$^{42}$ erg s$^{-1}$, that again lies between than the values from the model M5 and M3$^{\ddagger}$. 
The high resolution mid-IR core fluxes/luminosities presented in \cite{Gandhi09} are reported to be least contaminated 
by the circumnuclear starburst and primarily represent the emission from the torus.
The above comparisons for the 2.0 - 10 keV intrinsic luminosities suggest that the covering factor lies between 
50$\%$ to 100$\%$.
Finally, the low ratio between the soft power law flux and the intrinsic emission ($<0.01$, Table~2) suggests a low scattering efficiency, 
which again, may be due to a high covering factor of the thick absorber, leaving only a small fraction of the solid angle free for the primary 
continuum to reach the warm scatterer. 
A similar low flux ratio {\bf ($<0.01$)} of the unabsorbed powerlaw to the intrinsic 2.0 - 10 keV emission is  obtained with the 
model `M5' which considers an axi-symmetric toroidal geometry of the reprocessing material around the AGN.
The actual scattered fraction may be even lower, given that part of the soft power law may be 
due to the starburst emission.
\section{Conclusions}
We analyzed the {\em Suzaku} XIS + HXD-PIN observation of the AGN in NGC~5135. We found that the 0.5~-~50~keV spectrum can be 
reproduced by a model consisting of: (i) a soft component, characterized by a thermal plasma model (the {\tt MEKAL}) 
of temperature kT $\sim$~0.70~keV plus an unabsorbed  steep power law ($\Gamma$ $\sim$ 2.7), (ii) a hard component best fitted by 
an absorbed power law (N$_{\rm H}$ $\sim$ 2.5 $\times$ 10$^{24}$ cm$^{-2}$, $\Gamma$ $\sim$ 1.6), a reflection continuum
component and a prominent Fe K$\alpha$ line with EW~$\sim$~2.1~keV. 
{\em Suzaku} broad-band energy coverage allowed us to accurately measure the absorbing column density, intrinsic AGN luminosity and the 
reflection component. We attempted various physically motivated models and all the models confirm that the AGN in NGC 5135 is obscured by 
Compton-thick material. The {\tt MYTORUS} model which considers the geometry of the reprocessing material as a azimuthally symmetric torus 
suggests that the torus is viewed nearly edge on.
Both the estimates on the reflection strength and the comparison between the emission line flux and the continuum flux
suggest that the obscuring torus may cover 90$\%$ of the line-of-sight.
Our {\em Suzaku} X-ray spectral study of NGC~5135 may represent a case study of some of the complex features in  
Compton-thick obscured AGNs.
\section*{Acknowledgments}
This research has made use of data obtained from the {\em Suzaku} satellite, 
a collaborative mission between the space agencies of Japan (JAXA) and the USA (NASA).
Authors also thank to the anonymous referee for useful comments that helped to improve the quality of the paper.
\bibliographystyle{mn2e}
\bibliography{my}

\bsp

\label{lastpage}

\end{document}